\documentclass{IEEEoj-data}
\usepackage{cite}
\usepackage[utf8]{inputenc}
\usepackage{amsmath,amssymb,amsfonts}
\usepackage{algorithmic}
\usepackage{graphicx,color}
\usepackage{textcomp}
\usepackage[hyphens]{url}
\usepackage{dirtree}
\usepackage{hyperref}
\usepackage{balance}
\usepackage{url}
\usepackage{float}
\usepackage{array}
\def\BibTeX{{\rm B\kern-.05em{\sc i\kern-.025em b}\kern-.08em
    T\kern-.1667em\lower.7ex\hbox{E}\kern-.125emX}}
\AtBeginDocument{\definecolor{ojcolor}{cmyk}{0.93,0.59,0.15,0.02}}

\begin{document}

\title{\textcolor{black}{Descriptor:} \textcolor{ieeedata}{\textit{Extended-Length Audio Dataset for Synthetic Voice Detection and Speaker Recognition}} (ELAD-SVDSR)}

\author{Rahul Vijaykumar\authorrefmark{1}, Ajan Ahmed\authorrefmark{1}, John Parker\authorrefmark{1}, Dinesh Pendyala\authorrefmark{1}, Aidan Collins\authorrefmark{1}, {Stephanie Schuckers}\authorrefmark{1} and Masudul H. Imtiaz\authorrefmark{1}}

\affil{Dept of Electrical and Computer Engineering, Clarkson University, Potsdam, NY, USA.}

\corresp{CORRESPONDING AUTHOR: Masudul H. Imtiaz (e-mail: mimtiaz@clarkson.edu).}

\markboth{IEEE-DATA  Descriptor Article Template}{Author \textit{et al.}}

\begin{abstract}
This paper introduces the Extended-Length Audio Dataset for Synthetic Voice Detection and Speaker Recognition (ELAD-SVDSR), a resource specifically designed to facilitate the creation of high-quality deepfakes and support the development of detection systems trained against them. The dataset comprises 45-minute audio recordings from 36 participants, each reading various newspaper articles recorded under controlled conditions and captured via five microphones of differing quality. By focusing on extended-duration audio, ELAD-SVDSR captures a richer range of speech attributes—such as pitch contours, intonation patterns, and nuanced delivery—enabling models to generate more realistic and coherent synthetic voices. In turn, this approach allows for the creation of robust deepfakes that can serve as challenging examples in datasets used to train and evaluate synthetic voice detection methods. As part of this effort, 20 deepfake voices have already been created and added to the dataset to showcase its potential. Anonymized metadata accompanies the dataset on speaker demographics. ELAD-SVDSR is expected to spur significant advancements in audio forensics, biometric security, and voice authentication systems.\\
 \\ 
 \\ 
{\textcolor{ieeedata}{\abstractheadfont\bfseries{IEEE SOCIETY/COUNCIL}}} Signal Processing Society (SPS)\\  
 \\
 {\textcolor{ieeedata}{\abstractheadfont\bfseries{DATA DOI/PID}}} 10.21227/ab5w-0c23\\ 
  
 {\textcolor{ieeedata}{\abstractheadfont\bfseries{DATA TYPE/LOCATION}}}   Audio; Potsdam, NY, USA

\end{abstract}


\maketitle

\section*{BACKGROUND} 
The rapid advancement of text-to-speech (TTS) and deep learning techniques has enabled the production of highly realistic synthetic voices, often referred to as deepfakes \cite{van2016wavenet}. Early TTS systems were limited in quality primarily due to short-duration training data, which provided only a narrow sampling of speech features. However, modern approaches like WaveNet \cite{van2016wavenet} and Tacotron 2 \cite{shen2018natural} allow for longer length audio input and contain more sophisticated architectures.

As the fidelity of TTS models improves, concerns about misuse for impersonation and fraud have grown \cite{kinnunen2017asvspoof}. Moreover, detecting synthetic audio has become increasingly challenging, leading to dedicated efforts such as the ASVspoof initiative \cite{todisco2019asvspoof} to develop and benchmark anti-spoofing methods. While many detection algorithms perform well on short speech segments, longer-duration recordings can better reveal subtle artifacts in synthesis\cite{arik2017deep}.

Extended-length audio also benefits speaker recognition systems, which rely on robust feature representations extracted from diverse acoustic conditions \cite{kinnunen2010overview}. Longer recordings allow for more comprehensive modeling of a speaker’s unique vocal traits, helping TTS models produce stronger deepfake voices and providing richer data for detection systems to identify the presence of synthetic elements. Despite these advantages, there remains a shortage of publicly available datasets that provide extended-duration recordings suitable for both training and evaluating deepfake generation and detection algorithms. The dataset presented in this paper, ELAD-SVDSR, addresses this gap by offering 45-minute recordings from 36 participants, alongside 20 deepfake samples generated.

\subsection*{Related Work}
The VCTK Corpus \cite{veaux2017cstr} is a multi-speaker English dataset recorded under controlled studio conditions. It contains roughly 110 speakers, each providing short read passages. Although this dataset has good speaker diversity, it focuses on shorter utterances and lacks synthetic parallel data.

LibriSpeech \cite{panayotov2015librispeech} is a large corpus derived from public domain audiobooks, offering around 1,000 hours of speech from more than 2,400 speakers. While it provides extended recordings for ASR, it lacks controlled microphone variations and does not include high-fidelity synthetic samples.

VoxCeleb \cite{nagrani2017voxceleb} consists of thousands of speakers drawn from YouTube interviews. Its real-world noise and broad speaker coverage are strengths for speaker identification tasks, but the dataset primarily contains shorter clips, lacks microphone diversity, and does not include synthetic utterances.

LJSpeech \cite{ljspeech17} is a single-speaker dataset with about 24 hours of read speech. It is widely used for building TTS models due to its consistent recording environment. However, the single-speaker limitation and relatively short utterances reduce its suitability for multi-speaker deepfake research.

Mozilla’s Common Voice \cite{ardila2020commonvoice} is a crowdsourced dataset with contributions from global volunteers. It supports multiple languages and speaker demographics but varies significantly in recording quality and clip length, and it does not offer paired synthetic samples for deepfake detection.

TIMIT \cite{garofolo1993timit} is a classic dataset featuring phonetically balanced, short utterances in English. With high-quality, time-aligned transcriptions, TIMIT remains popular for phonetic research. However, the limited speaker set and short, lab-recorded prompts render it insufficient for extended deepfake generation and detection.

AISHELL-1 \cite{bu2017aishell} is an open-source Mandarin speech corpus recorded under relatively quiet conditions. While it provides useful data for non-English speech research, it consists mostly of short utterances and lacks any synthetic component, limiting its value for deepfake-related studies.

Overall, these datasets (summarized in Table~\ref{tab:dataset_comparison}) have significantly advanced speech technology but do not comprehensively address the need for extended-duration recordings, controlled microphone diversity, and integrated synthetic samples. ELAD-SVDSR fills this gap by providing 45-minute recordings per speaker under multiple microphone conditions alongside high-quality deepfake audio.

\begin{table*}[h]
\centering
\caption{Comparison of Popular Speech Datasets with ELAD-SVDSR.}
\label{tab:dataset_comparison}
\begin{tabular}{|p{2.2cm}<{\centering}|p{1.4cm}<{\centering}|p{2.0cm}<{\centering}|p{2.5cm}<{\centering}|p{1.6cm}<{\centering}|p{1.6cm}<{\centering}|p{2cm}<{\centering}|}
\hline
\textbf{Dataset} & \textbf{\# Speakers} & \textbf{Total Hours} & \textbf{Avg.\ Clip Length} & \textbf{Synthetic Data?} & \textbf{Mic Diversity} & \textbf{Extended Per Speaker}\\
\hline

\textbf{VCTK \cite{veaux2017cstr}}      
 & $\sim110$  
 & $\sim44$    
 & Short (a few seconds) 
 & No         
 & Single, Studio
 & No  \\
\hline

\textbf{LibriSpeech \cite{panayotov2015librispeech}} 
 & 2,484+ 
 & $\sim1000$ 
 & Mostly a few seconds 
 & No         
 & Not specified 
 & Partial  \\
\hline

\textbf{VoxCeleb \cite{nagrani2017voxceleb}} 
 & 7,000+      
 & $\sim2000$ 
 & Short to medium ($<$10\,s) 
 & No         
 & In-the-wild
 & No  \\
\hline

\textbf{LJSpeech \cite{ljspeech17}}      
 & 1          
 & $\sim24$    
 & 10--15\,s 
 & No         
 & Single, Studio
 & No  \\
\hline

\textbf{Common Voice \cite{ardila2020commonvoice}} 
 & Thousands
 & Several thousand 
 & Mostly a few seconds 
 & No 
 & User-recorded, variable
 & No \\
\hline

\textbf{TIMIT \cite{garofolo1993timit}}
 & 630
 & $\sim5$ 
 & 3--4\,s
 & No
 & Single, Lab
 & No \\
\hline

\textbf{AISHELL-1 \cite{bu2017aishell}}
 & 400+
 & $\sim178$
 & Short utterances
 & No
 & Single, Quiet
 & No \\
\hline

\textbf{ELAD-SVDSR (Proposed)} 
 & 36         
 & $\sim27$  
 & 45\,min/speaker 
 & Yes (20 deepfakes)   
 & 5 distinct mics
 & Yes \\
\hline

\end{tabular}
\end{table*}

\section*{COLLECTION METHODS AND DESIGN} 

\subsection*{Institutional Review Board (IRB) Approval}
ELAD-SVDSR was developed following rigorous ethical guidelines and procedures approved by the Institutional Review Board (IRB Approval No. 24-42) at Clarkson University \cite{clarkson_irb}. This approval ensures that all aspects of the research involving human subjects adhere to the highest ethical standards, particularly regarding informed consent, data confidentiality, and the overall treatment of participants.

\paragraph{Ethical Considerations and Informed Consent}
The IRB’s primary role is to safeguard the rights and well-being of research participants. All participants were fully informed about the study's nature, potential risks and benefits, and their rights as participants. Each participant signed an informed consent form detailing the study's purpose, procedures, data-sharing permissions, and measures to ensure their confidentiality. Flyers were distributed throughout the university campus to recruit participants.

\paragraph{Data Confidentiality and Security}
The IRB addressed the critical concern of protecting participant data. All voice recordings and associated metadata were anonymized before inclusion in the dataset, meaning no identifying information was linked to the tapes. After anonymization, all identifiable data were permanently deleted, and the consent forms were physically and securely stored at the university. The data was stored in secure, password-protected environments, accessible only to authorized researchers.

\subsection*{Participant Recruitment and Consent}
Participants for the ELAD-SVDSR dataset were recruited from the Clarkson University community through flyers and electronic communications. All potential participants were provided with detailed information about the study, including its objectives, procedures, and their rights as participants. Informed consent was obtained from all participants before the data collection began. 

\subsection*{Recording Environments}

\begin{figure}[h]
    \centering
   \includegraphics[width=1\linewidth]{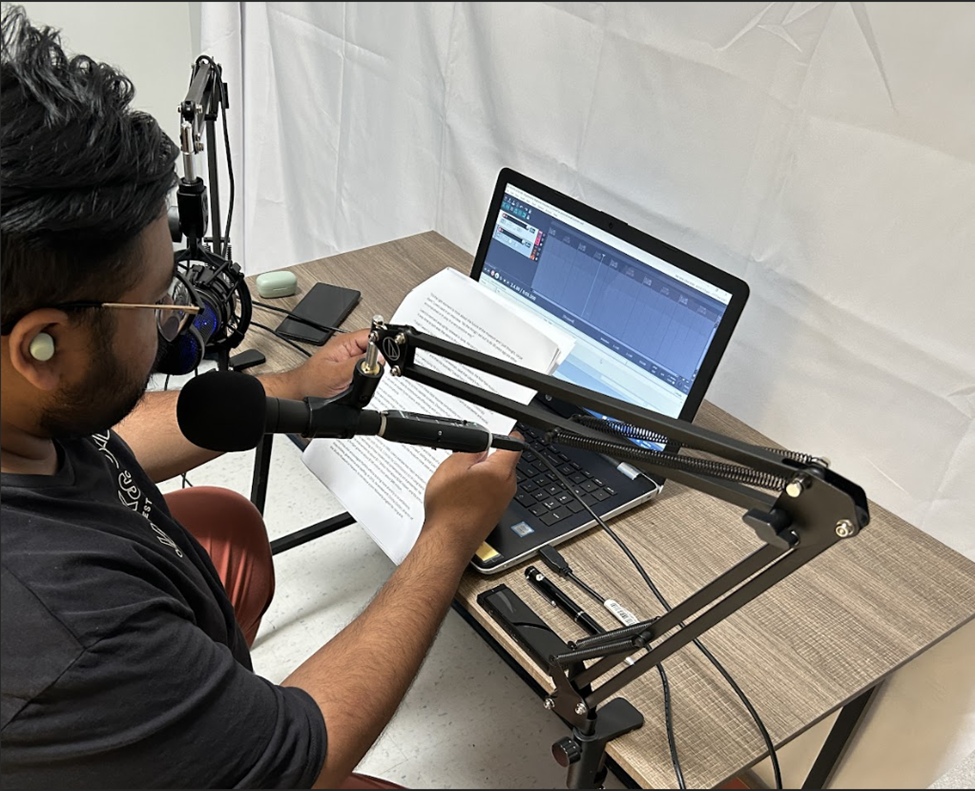}
    \caption{Live data collection Process}
    \label{fig:live_collection}
\end{figure}

The recording sessions were conducted in a closed-room environment with minimal external noise and maintaining consistent acoustic conditions. Participants were seated as close as possible to all recording equipment. Additionally, researchers followed a uniform protocol during all sessions—monitoring microphone placement, participant comfort, and other environmental factors—to ensure high-quality and reliable audio samples. Figure~\ref{fig:live_collection} shows the environment of live data collection process.

\subsection*{Recording Procedure}
Participants were instructed to read aloud three contemporary news articles during their recording sessions making the data text-dependent. The first article, sourced from The New York Times, covered economic and political topics related to tariff policies \cite{nyt_article}. The remaining two were taken from TIME magazine: one discussing issues surrounding pregnancy criminalization in a post-Dobbs context \cite{time_pregnancy} and another examining how scams and fraud schemes evolve in the digital age \cite{time_scams}. These articles were specifically chosen due to their diverse vocabulary and subject matter, which help cover a wide range of phoneme variations in the English language to reflect distinct intonations, linguistic structures, and contexts. 

\subsection*{Recording Equipment}

All voice recordings for this study were captured using:
\begin{itemize}
    \item \textbf{Audio-Technica AT2020} \cite{audio-technica2020}
    \item \textbf{Shure SM58} \cite{shure_sm58}
    \item \textbf{TOZO A1} \cite{tozoA1}
    \item \textbf{Inni Oasis R1} \cite{inniOasisR1}
    \item \textbf{ZIPCIDE Digital Voice Activated Recorder (Spy Pen)} \cite{zipcidePen}
\end{itemize}

\begin{figure}[h]
    \centering
    \includegraphics[width=0.2\textwidth]{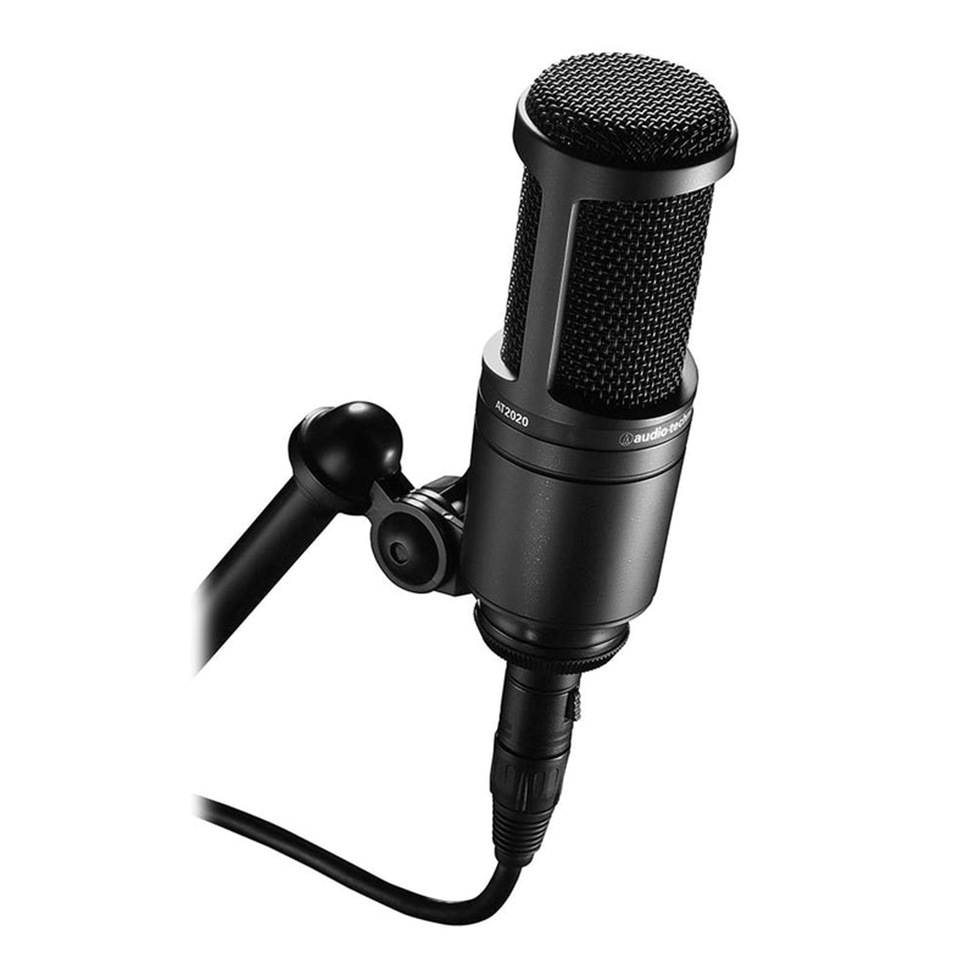}\\
    \includegraphics[width=0.2\textwidth]{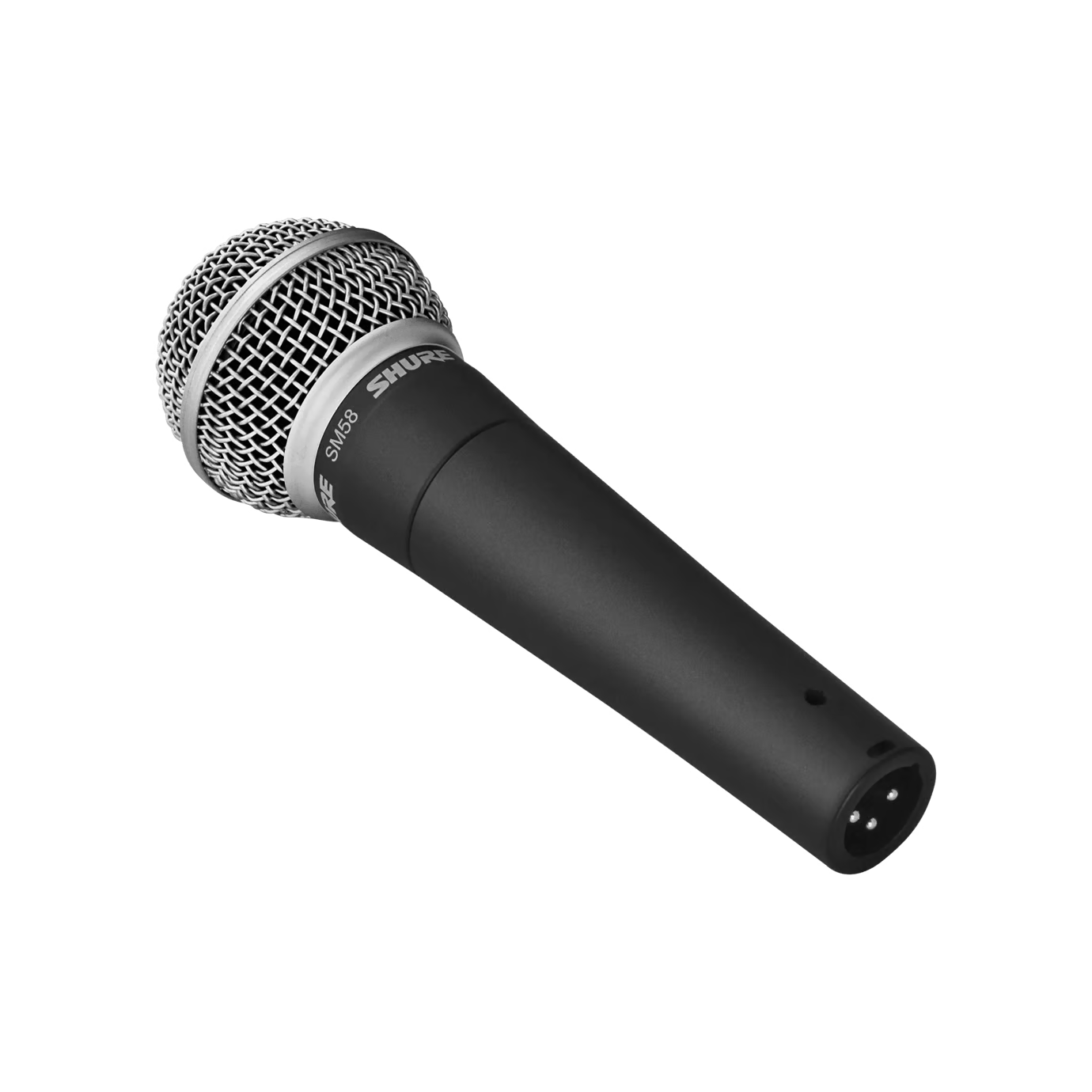}\\
    \includegraphics[width=0.2\textwidth]{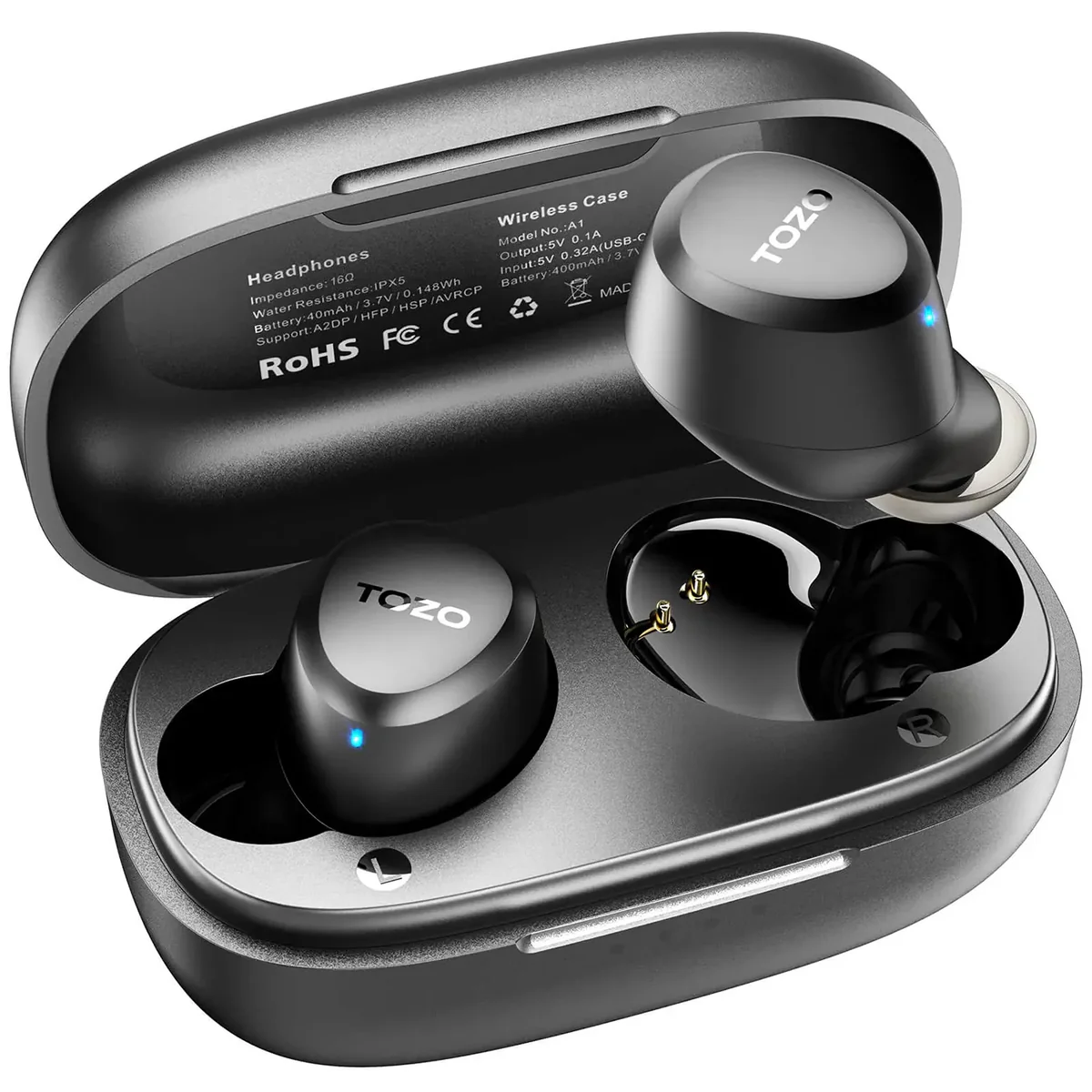}\\
    \includegraphics[width=0.2\textwidth]{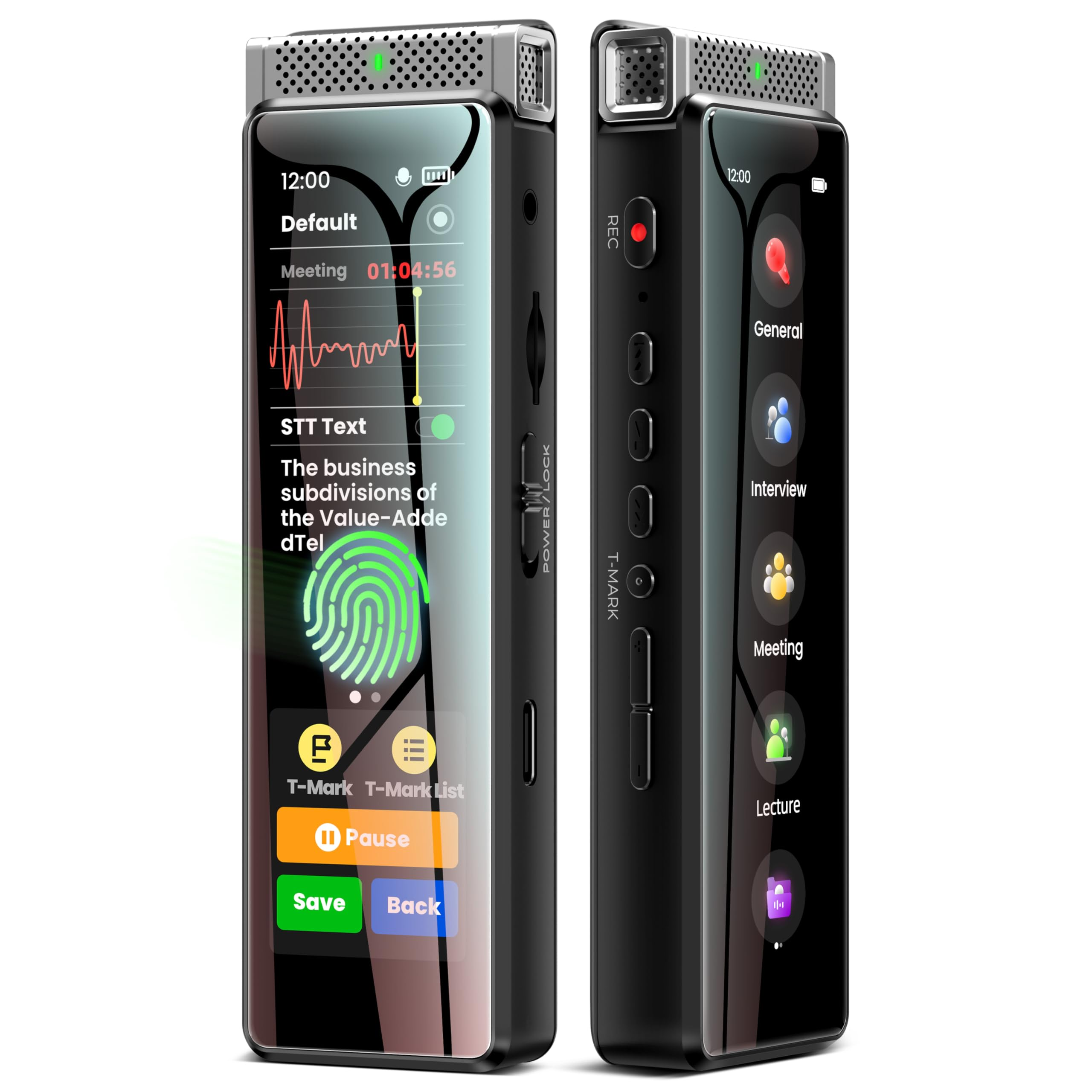}\\
    \includegraphics[width=0.2\textwidth]{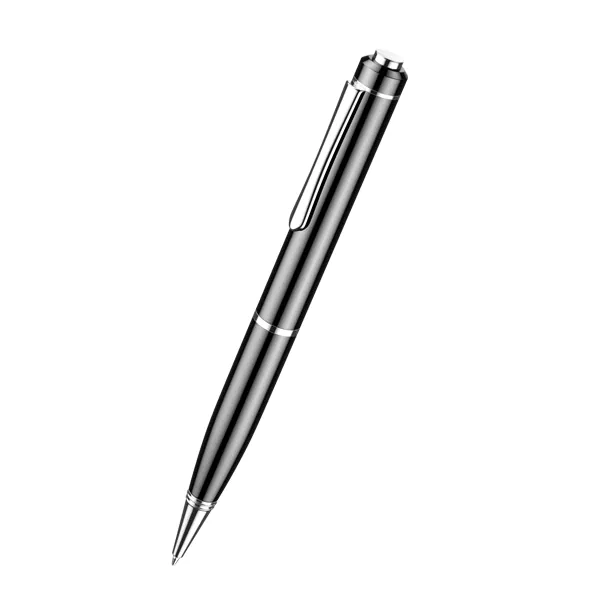}
    \caption{Microphones and recording devices used in this study (from top to bottom): 
    Audio-Technica AT2020, 
    Shure SM58, 
    TOZO A1 earbuds, 
    Inni Oasis R1 recorder, 
    and the ZIPCIDE Spy Pen.}
    \label{fig:mics}
\end{figure}

Each device’s complete specifications are publicly available through its respective manufacturer’s website \cite{audio-technica2020, shure_sm58, tozoA1, inniOasisR1, zipcidePen}. The recordings were made at a standard sampling rate of 44.1 kHz. The Audio-Technica AT2020 operates within a frequency range of 20 Hz to 20 kHz, has a sensitivity of -37~dB, and can handle a maximum sound pressure level (SPL) of 144~dB. It requires 48V phantom power to operate and connects via an XLR output\cite{audio-technica2020}. The Shure SM58 operates within a frequency response range of 50 Hz to 15 kHz and is optimized to emphasize clarity in the vocal midrange while attenuating low-frequency background noise\cite{shure_sm58}. The TOZO A1 earbuds capture near-field audio through an integrated microphone and connect via Bluetooth \cite{tozoA1}. The Inni Oasis R1 is a touchscreen digital recorder capable of high-fidelity audio capture under relatively quiet conditions \cite{inniOasisR1}. Finally, the ZIPCIDE Digital Voice Activated Recorder (commonly referred to as a “spy pen”) discreetly captures speech, functioning in a compact form factor for on-the-go recording \cite{zipcidePen}. Figure~\ref{fig:mics} displays the microphones used for data collection.

\section*{VALIDATION AND QUALITY} 
\subsection*{Data Quality Control}

Each recording was manually reviewed for clarity and consistency to ensure the highest quality data. The final dataset excluded records that contained any clipping, inconsistent sampling rates (44.1 kHz), unclear or interrupted speech, and technical malfunctions.

\subsection*{Speaker Diversity}

As shown in Table~\ref{tab:diversity}, broad representation across age, gender, and accents is ensured.

\begin{table}[h]
\centering
\caption{Participant Demographic Information}
\begin{tabular}{|>{\centering\arraybackslash}m{1.5cm}|>{\centering\arraybackslash}m{2.5cm}|>{\centering\arraybackslash}m{1.5cm}|}
\hline
\textbf{Category}           & \textbf{Subcategory}               & \textbf{Number of Participants} \\
\hline
\textbf{Gender}              & Man                               & 26 \\
                             & Woman                             & 10 \\
                             & Prefer not to respond             &  0\\
\hline
\textbf{Race}                & Caucasian                         & 7 \\
                             & Black                             & 5\\
                             & Hispanic                          & 1 \\
                             & Native American                   & 1 \\
                             & Middle Eastern                    & 1 \\
                             & Indian                            & 10 \\
                             & Asian                             & 10 \\
                             & Other                             & 1 \\
\hline
\textbf{Age}                 & 18-25 years                       & 15\\
                             & 26-30 years                       & 16\\
                             & 31-40 years                       & 5 \\
\hline
\end{tabular}
\label{tab:diversity}
\end{table}

\subsection*{Audio Quality Metrics}
The audio properties highlighted in Table~\ref{tab:audio_quality} demonstrate the key attributes. Noise levels were measured in A-weighted decibels (dBA) using calibrated microphones. The noise levels ranged between 2.66 dBA and 10.82 dBA, with an average of 6.27 dBA. This range reflects noise close to near-silent environments (4.86~dB)\cite{audicus, lexiehearing, noiseawareness}. 

\begin{table}[h]
    \centering
    \caption{Summary of Audio properties in the ELAD-SVDSR Dataset}
    \label{tab:audio_quality}
    \begin{tabular}{|p{4.5cm}|c|}
        \hline
        \textbf{Metric} & \textbf{Value} \\ \hline
        \textbf{Sampling Rate} & 44.1 kHz \\ \hline
        \textbf{Bit Depth} & 16-bit \\ \hline
        \textbf{Average Noise Level of all audio files} & 6.27~dB \\ \hline
        \textbf{Total Duration of Recordings} & 27 hours \\ \hline
    \end{tabular}
\end{table}

Table~\ref{tab:snr_values} shows the notably high signal-to-noise ratio (SNR) values obtained from this analysis. The SNR is computed using the formula as Ahmed et al.\cite{ahmed2024descriptor} as:
\[
\text{SNR (in dB)} = 10 \times \log_{10} \biggl( \frac{P_{\text{signal}}}{P_{\text{noise}}} \biggr),
\]
where \(P_{\text{signal}}\) and \(P_{\text{noise}}\) are the mean-square power of the speech signal and the background noise, respectively. Specifically,
\[
P = \frac{1}{N} \sum_{n=1}^{N} x[n]^2,
\]
with \(x[n]\) denoting the amplitude of the signal or noise at the \(n\)-th sample, and \(N\) is the total sample count.

To approximate the noise level, any audio segment whose amplitude falls below a set threshold—derived from the average overall signal energy—was designated as noise. The remaining portion of the audio was treated as the speech signal. Two additional SNR measures were evaluated: Segmented SNR (SegSNR) and Frequency-Weighted SNR (fwSNR).

\textbf{Segmented SNR (SegSNR):}
This metric divides the audio into short, fixed-length frames (around 20--30\,ms). The SNR for each segment is calculated as:
\[
\text{SegSNR (in dB)} = 10 \times \log_{10} \biggl(\frac{P_{\text{segment}}}{P_{\text{noise}}}\biggr),
\]
where \(P_{\text{segment}}\) corresponds to the mean-square power of that segment, and \(P_{\text{noise}}\) is the noise power. The final SegSNR is obtained by averaging across all segments, offering a granular view of time-varying noise levels.

\textbf{Frequency-Weighted SNR (fwSNR):}
An A-weighting filter is applied to emphasize frequencies crucial for human hearing (typically 500\,Hz--5\,kHz). The fwSNR is then given by:
\[
\text{fwSNR (in dB)} = 10 \times \log_{10} \biggl(\frac{P_{\text{A-weighted signal}}}{P_{\text{A-weighted noise}}}\biggr),
\]
where speech and noise powers are computed after the filter is applied. As with the other methods, noise segments are estimated based on low-energy detection. The scripts for SNR calculations can be accessed at:
\url{https://github.com/ahmedajan/SNR_Calculation_For_VPQAD/tree/main/VPQAD_SNR}

\begin{table}[ht]
\centering
\caption{Summary of Different SNR Metrics for ELAD-SVDSR }
\begin{tabular}{|>{\centering\arraybackslash}m{2cm}|
                >{\centering\arraybackslash}m{1cm}|
                >{\centering\arraybackslash}m{1cm}|
                >{\centering\arraybackslash}m{1cm}|}
\hline
\textbf{Metric}         & \textbf{Highest Value (dB)} & \textbf{Lowest Value (dB)} & \textbf{Mean Value (dB)} \\
\hline
SNR                     & 68.12                       & 38.60                      & 57.41                    \\
\hline
Segmented SNR (SegSNR)  & 55.55                       & 35.23                      & 54.16                    \\
\hline
Frequency-Weighted SNR  & 69.90                       & 40.01                      & 58.72                    \\
\hline
\end{tabular}
\label{tab:snr_values}
\end{table}

Under these high-SNR conditions, background noise remains minimal relative to the speech signal, indicating a more favorable environment for intelligibility. 

\subsection*{Generation of DeepFakes}

A total of 20 deepfake voices were generated using Tortoise TTS \cite{tortoiseTTS}. Before feeding each participant’s 45-minute speech into the model, the data was manually pre-processed using Audacity audio software\cite{audacity2023}. The data was manually heard, silences were trimmed, and disfluencies (fillers, long pauses)were removed; this manually annotated audio was segmented into $\sim$10-second clips. Two of these clips were separated randomly to compare with the deepfakes generated. The other clips were fed into the model for training. Deepfakes were generated such that the synthetic audio had the same sentence as the two test clips separated for proper matching. 

\paragraph{VeriSpeak Match Scores}
To evaluate the quality of deepfakes generated, VeriSpeak\cite{verispeakTool}, an industry-standard speaker recognition performance measurement tool  was used. VeriSpeak Match Scores quantified how closely the generated voice resembles the genuine speaker’s characteristics. Although VeriSpeak internally calculates match scores on an absolute scale, we present \emph{normalized} percentages based on the original-to-original comparison as 100\%. Table~\ref{tab:verispeak_scores} summarizes the results for 3 random samples of these normalized results drawn from the dataset.

\begin{table}[ht]
\centering
\caption{Normalized VeriSpeak Match Scores for Original vs.\ Deepfake Recordings}
\label{tab:verispeak_scores}
\begin{tabular}{@{}lccc@{}}
\hline
\textbf{Participant} & \textbf{Recording Type} & \textbf{Match Score (Normalized \%)} \\
\hline
Subject 3 & Original   & 100.0 \\
    & Deepfake   &  41.7 \\
\hline
Subject 6 & Original   & 100.0 \\
    & Deepfake   &  39.3 \\
\hline
Subject 18 & Original   & 100.0 \\
    & Deepfake   &  34.5 \\
\hline
\end{tabular}
\end{table}

While the similarity percentages may seem low, they are of good quality when it comes to generation of deepfakes as demonstrated by Table~\ref{tab:dataset_similarity}. For this comparison, 10 different audio samples are taken from each of the different datasets to generate deepfakes and verySpeak match scores. The DeepFake-to-original scores are then normalized as a percentage against the Original-to-Original scores and the results are shown.

\begin{table}[ht]
\centering
\caption{Mean Percentage Normalized Similarity Scores for Various Datasets from the Literature Review}
\label{tab:dataset_similarity}
\begin{tabular}{@{}lc@{}}
\hline
\textbf{Dataset}           & \textbf{Mean Similarity (\%)} \\
\hline
VCTK \cite{veaux2017cstr}  & 15.6 \\
LibriSpeech \cite{panayotov2015librispeech}  & 18.9 \\
VoxCeleb \cite{nagrani2017voxceleb}          & 24.2 \\
LJSpeech \cite{ljspeech17}                  & 28.9 \\
Common Voice \cite{ardila2020commonvoice}    & 19.3 \\
TIMIT \cite{garofolo1993timit}               & 10.1 \\
AISHELL-1 \cite{bu2017aishell}               & 2.4 \\
\textbf{ELAD-SVDSR (Proposed)}               & \textbf{37.2} \\
\hline
\end{tabular}
\end{table}

\section*{RECORDS AND STORAGE}

\subsection*{Data Processing and Storage}
After each recording session, the data was stored in a secure, password-protected digital archive, with access only to authorized researchers. This was done to maintain privacy while all identifying information was separated from the audio files and permanently deleted from the recordings. The files were anonymized by assigning each recording a unique identification code, ensuring no personal information was linked to the voice data. This ensured that the dataset could now be made public while maintaining participants' privacy according to the IRB regulations.

\subsection*{File Naming Conventions and Folder Architecture}
\begin{figure}
    \centering
    \includegraphics[width=0.8\linewidth]{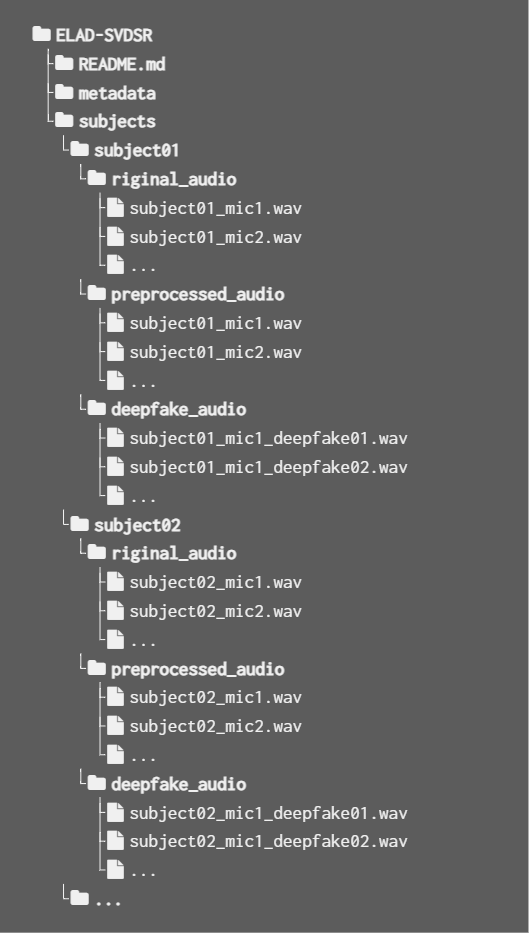}
    \caption{Directory Architecture for ELAD-SVDSR}
    \label{fig:elad_arch}
\end{figure}

At the top level, the dataset contains essential files such as the README (explaining the usage of the dataset, especially the mic numbers and their corresponding microphones) and a metadata folder that holds CSV files with details on the speaker demographics. The main body of the dataset is contained within the "subjects" directory, where each subject is assigned its folder (e.g., subject01, subject02, etc.). Within each subject folder, there are three subdirectories: "original\_audio," which stores the raw recordings; "preprocessed\_audio," which contains annotated versions of these recordings; and "deepfake\_audio," which holds the synthesized deepfake outputs. The files are all in .wav* format and the naming convention indicates the origin of each audio file; for instance, a file named "subject01\_mic1.wav" in the original\_audio folder signifies a recording from subject 01 captured with Shure SM58, while "subject01\_mic1\_deepfake01.wav" in the deepfake\_audio folder denotes the first deepfake generated from that same microphone recording as shown in Figure~\ref{fig:elad_arch}.

\balance

\section*{INSIGHTS AND NOTES} 
\subsection*{Accessing data}
The Extended-Length Audio Dataset for Synthetic Voice Detection and Speaker Recognition(ELAD-SVDSR) is intended exclusively for academic research. To obtain access, researchers are required to sign the End User License Agreement (EULA), which can be requested via email at \href{mailto:mimtiaz@clarkson.edu}{mimtiaz@clarkson.edu} or downloaded directly from the IEEE Dataport. A signed EULA must then be returned to this email address. Only emails originating from academic accounts will be accepted.

\subsection*{Dataset Limitations}
ELAD-SVDSR’s extended-duration recordings capture a range of speech characteristics, yet several practical constraints remain. Although the dataset features 36 speakers, this may not fully reflect the linguistic diversity seen in broader populations. The collection was conducted in a controlled, low-noise environment, producing high-fidelity audio that may not mirror real-world acoustic conditions. Additionally, the focus on English newspaper readings restricts the applicability of the dataset to other languages and more spontaneous speech tasks. The deepfake profiles were generated using a specific text-to-speech model, which can introduce biases that limit their generalizability. Future expansions to include more varied recording scenarios, a broader speaker population, and alternative speech domains would help mitigate these limitations and enhance ELAD-SVDSR’s utility.

\section*{SOURCE CODE AND SCRIPTS} 

The scripts for calculating signal-to-noise ratios (SNR), segmenting audio files, and evaluating speech quality metrics are publicly accessible. These are available via GitHub and the repository link for scripts used in this dataset are available at: \url{https://github.com/ahmedajan/SNR_Calculation_For_VPQAD/tree/main/VPQAD_SNR}

\section*{ACKNOWLEDGEMENTS AND INTERESTS}
R.V., A.A., J.P. and D.P. conducted the data collection. J.P., D.P. and A.C. conducted preprocessing of the data. A.A. curated the data collection, analyzed the data and wrote the manuscript. S.S. and M.H.I. reviewed the data collection, curation, and analysis. All authors reviewed the manuscript.

This work is supported by the Center for Identification Technology Research and the National Science Foundation under Grant No. 1650503.

The article authors have declared no conflicts of interest.
\bibliographystyle{IEEEtran}
\bibliography{references}
\end{document}